\documentclass [submission,copyright,creativecommons]{eptcs}
\providecommand{\event}{Workshop on Adaptable Cloud Architectures\\
DisCoTec 2025 - 20th International Federated Conference\\
on Distributed Computing Techniques} 

\providecommand{\thisvolume}[1]{this volume of EPTCS, Open Publishing Association}

\ifpdf{}
  \usepackage{underscore}         
  \usepackage[T1]{fontenc}        
\else
  \usepackage{breakurl}           
\fi

\usepackage{iftex}
\usepackage[dvipdfm]{graphicx}
\usepackage{bmpsize}
\usepackage[english]{babel}
\usepackage{wrapfig}
\usepackage{lscape}
\usepackage{rotating}
\usepackage{listings}
\usepackage{amsfonts}
\usepackage{array}
\usepackage{booktabs}
\usepackage{amsmath}
\usepackage[version=4]{mhchem}
\usepackage{siunitx}
\usepackage{longtable,tabularx}
\usepackage{times}
\usepackage{subcaption}
\usepackage{xcolor}
\usepackage{titlesec}
\usepackage{multirow}
\usepackage{enumitem}

\usepackage{calc}              %

\newlength{\tabwidthi}
\newlength{\tabwidthii}
\newlength{\tabwidthiii}
\newcommand{\tabsize}{\footnotesize}
\newcommand{\thead}[1]{\tabsize\textbf{#1}}

\usepackage{hyperref}

\definecolor{codegreen}{rgb}{0,0.6,0}
\definecolor{codegray}{rgb}{0.5,0.5,0.5}
\definecolor{codepurple}{rgb}{0.58,0,0.82}
\definecolor{backcolour}{rgb}{0.95,0.95,0.92}
\lstdefinestyle{mystyle}{
    backgroundcolor=\color{backcolour},
    commentstyle=\color{codegreen},
    keywordstyle=\color{magenta},
    numberstyle=\tiny\color{codegray},
    stringstyle=\color{codepurple},
    basicstyle=\ttfamily\footnotesize,
    breakatwhitespace=false,
    breaklines=true,
    keepspaces=true,
    numbers=left,
    numbersep=4pt,
    showspaces=false,
    showstringspaces=false,
    showtabs=false,
    tabsize=2,
    language={c++},
    classoffset=1, 
    alsoletter={-,>},
    morekeywords={->, SYSTEM, physical, logical, pure, dt, int, float, init, then, link, to, byte, mergeplug, splitplug},
    keywordstyle=\color{codepurple},
    classoffset=0
}
\title{Fancy Some Chips for Your TeaStore? \\
Modeling the Control of an Adaptable Discrete System
\thanks{This work was supported by the ANR grant ANR-23-CE25-0004 (ADAPT). O.~Kouchnarenko was supported by the EIPHI Graduate School (grant number ANR-17-EURE-0002).}}

\author{
Anna~Gallone
\institute{
Université Marie et Louis Pasteur, CNRS UMR6174, Institut FEMTO-ST, F-25000 Besançon, France}
\email{Anna.Gallone@femto-st.fr}
\and 
Simon~Bliudze
\institute{Univ. Lille, Inria, CNRS, Centrale Lille, UMR 9189 CRIStAL, F-59000 Lille, France}
\email{Simon.Bliudze@inria.fr}
\and
Sophie~Cerf
\institute{Univ. Lille, Inria, CNRS, Centrale Lille, UMR 9189 CRIStAL, F-59000 Lille, France}
\email{Sophie.Cerf@inria.fr}
\and
Olga~Kouchnarenko
\institute{
Université Marie et Louis Pasteur, CNRS UMR6174, Institut FEMTO-ST, F-25000 Besançon, France}
\email{Olga.Kouchnarenko@femto-st.fr}
}

\def\titlerunning{Chips: Modeling the Control of an Adaptable Discrete System}
\def\authorrunning{A.~Gallone et al.}
\begin{document}
\maketitle

\begin{abstract}
When designing new web applications, developers must cope with different kinds of constraints relative to the resources they rely on: software, hardware, network, online micro-services, or any combination of the mentioned entities. Together, these entities form a complex system of communicating interdependent processes, physical or logical. It is very desirable that such system ensures its robustness to provide a good quality of service. In this paper we introduce Chips, a language that aims at facilitating the design of models made of various entwined components. It allows the description of applications in the form of functional blocks. Chips mixes notions  from control theory and general purpose programming languages to generate robust component-based models. This paper presents how to use Chips to systematically design, model and analyse a complex system project, using a variation of the Adaptable TeaStore application as running example.
\end{abstract}

\section*{Resources}
A GitHub repository for the Chips language is available at the address \url{https://github.com/NwaitDev/Chips_Public}.\\
The source code of the Chips Adaptable TeaStore model and its BIP translation are available at \url{https://github.com/NwaitDev/TeaStore-Variation}.

\section{Introduction}
Chips, standing for Control of Hierarchical Interconnected Programmable Systems, is a language for designing, modeling and simulating distributed systems that execute tasks having strong interactions with their environment. Though its primary field of application is Cyber-Physical Systems (CPS), this paper proposes a Chips implementation of a model for a Cloud Computing (CC) system, the Adaptable TeaStore application~\cite{BDGLZZ25}. Adaptivity focus is put on the caching strategy: according to the response time of the website, the system changes the size of the cache at execution-time to maintain the service while reducing delays. To ensure an efficient adaptation, the followed methodology is grounded on Control Theory formalism as proposed by Filieri et al.~\cite{filieri-software-2015}. Chips mixes a synchronous aspect with syntactic constructs to easily configure and connect many components. Section~\ref{sec:ChipsOverview} gives an overview of our language's features. They allow to draw a full Model Based Development pipeline for an adaptable system project. Such pipeline is composed of the following steps (each being illustrated by its application on the Adaptable TeaStore example):
\begin{itemize}
  \item The design of an adaptable Chips model for the application, Section~\ref{sec:DesigningAdaptableTeaStore}.
  \item Its potential declination with the duplication of the system's components to balance workload or another parameter of interest, Section~\ref{sec:APTwist}.
  \item The transformation of the Chips model into a form that allows testing, simulation or code generation, Section~\ref{sec:ExpSetup}.
\end{itemize}
Section~\ref{sec:results} presents the results obtained when applied to the Adaptable TeaStore. Section~\ref{sec:relatedwork} discusses some common aspects and differences of our methodology with other works revolving around real-time adaptation of computing systems. Finally, some insight is given on what is intended as future work for a better adaptivity of models designed with Chips and for the language itself. Additional graphical representations of the concepts presented in this paper are provided in Appendix~\ref{secn:appendix}.

\section{Overview of the Chips language}
\label{sec:ChipsOverview}

Cloud Computing systems share many characteristics with structured distributed CPSs aimed by Chips, namely, the coordination of multiple devices over communication protocols, the repeating entry and exit of devices in the network or the need to maintain a service when a component is temporarily unavailable. Hence, Chips can be effectively used for the design of CC systems though it is focused on the design of CPSs' models, and more particularly \textit{complex systems}\footnote{Hereafter, the term ``complex system'' will be used to describe a set of processing units achieving a task altogether, with each unit potentially being implemented by another kind of system, which may be complex as well.}. The following subsections introduce, in a first place, the technologies and concepts on which Chips lies, then presents the main principles of the language.

\subsection{Chips technical and theoretical base}
In this paper, we propose a compilation from Chips into BIP~\cite{basu-rigorous-2011}, which opens the way for many additional model based operations. BIP (for Behavior Interaction Priority) is a framework for developing interacting automata based models. Its rigorous implementation of synchronicity mechanisms makes it possible to derive these models through correct-by-construction programs generation, and model exploration tools are provided alongside the BIP compiler to ensure the right behavior of the designed applications. When it comes to the adaptation of a complex system, one can distinguish two kinds of adaptation for programmable components: behavioral adaptation and structural reconfiguration~\cite{puviani-taxonomy-2013}.
The former is about modifying the data manipulated among the components for the system to take decisions, while the latter is about modifying the way components interact, by modifying the number of instances of subsystems in use or their connections. Behavioral adaptation can be achieved with the BIP framework pretty easily, while structural reconfiguration requires more work, hence the development of DR-BIP (Dynamic Reconfigurable BIP)~\cite{TR-2018-3}. Providing a language to automatize the generation of BIP models should allow achieving adaptation goals in a systematic way. In this context, Chips proposes to take a step back from the BIP models, and to add upon them another abstraction layer taking advantage of two main concepts:
\begin{itemize}
  \item The first one being \textit{Control Theory} (CT for short)~\cite{zuazua-control-2003}, a well known engineering domain, where each component is modeled by a function. Additional control components allow the system behavior to adapt at run-time. By assembling the functions together, CT turns a physical system into a differential equation. Such equation solutions are the different behaviors to adopt for the control components so the system realizes a task as expected. Though control theory is mainly applied to the engineering of continuous systems, it can find applications for discrete systems too~\cite{hellerstein-feedback-2004}.
  \item The second concept is \textit{Aggregate Programming} (AP for short), a programming paradigm implementing collective communication primitives that rely on field calculus~\cite{viroli-distributed-2019}. The role components take in the collective behavior of the system is computed on the go by the state of variables continuously updated on each device according to their position. In such systems, leadership is assumed by the field of the values shared in the space and not by a particular device, making applications developed according to the aggregate programming principles extremely resilient to structural changes.
\end{itemize}

Our goal with the Chips language is to fill the gap between these engineering concepts and their concrete implementation.

\subsection{Chips programming principles}

\begin{figure}[t]
\begin{minipage}[b]{0.45\columnwidth}
\begin{lstlisting}[
  basicstyle=\tiny, %or \small or \footnotesize etc.
  caption=Code sample of Chips simplified TeaStore model components,
  label=codesample,
  captionpos=b
]
// extract last bit value for validation
pure user_action_is_valid(int user_action_data)
  -> ( (user_action_data & 0x1) == 0x1 )

// all other bits are the qty of images to fetch
pure user_action_nb_imgs(int user_action_data)
  -> (user_action_data >> 1)

logical interpretation_module(int user_action) init {
  int nbr_imgs = 0;
} then {
  if(!user_action_is_valid(user_action)){
    nbr_imgs = 0;
  } else {
    nbr_imgs = user_action_nb_images(user_action);
  }
} -> (nbr_imgs)

logical image_provider(int nb_imgs_to_provide) init {
  int [] cache = empty_set;
  time_since_last_req = 0;
} then {
  for i in n_up_to(nb_imgs_to_provide) {
    img_to_fetch = rnd_img_id();
    if( !find(img_to_fetch,cache) ){
      // simulating db request
    }
    lru_update(cache, img_to_fetch);
  }
} -> (1)

import "server.json" as server;

physical server(int server_resources, int user_action)
  -> (server_resources, user_action)

/* ... other definitions ... */
\end{lstlisting}
\end{minipage}
\hfill
\begin{minipage}[b]{0.47\columnwidth}
\begin{lstlisting}[
  basicstyle=\tiny, %or \small or \footnotesize etc.
  caption=System description of Chips simplified TeaStore model,
  label=codesample2,
  captionpos=b
]
SYSTEM {
  user user_instance;

  server server_instance;
  interpretation_module im_instance;
  image_provider ip_instance1;
  image_provider ip_instance2;
  link im_instance to server_instance;
  link ip_instance1 to server_instance;
  link ip_instance2 to server_instance;

  
  im_instance.in(server_instance.out[1]);

  splitplug(im_instance.out, profiling_flineunction);

  ip_instance1.in(im_instance.out);
  ip_instance2.in(im_instance.out);

  mergeplug(ip_instance1.out, aggregating_function);
  mergeplug(ip_instance2.out, aggregating_function);

  server_instance.in(ip_instance1.out, user_instance.out);
  server_instance.in(ip_instance2.out, user_instance.out);

  user_instance.in(server_instance.out[0]);
}\end{lstlisting}
\end{minipage}
\end{figure}

\paragraph{A synchronous programming language} The closest to control theory programming paradigm one can find in the state of the art is synchronous programming. Languages implementing this paradigm aim at describing flows of data instead of variables, and they transform dataflows assuming some operations can be made instantaneously. Lustre~\cite{caspi-lustre-nodate} and one of its derived language Heptagon~\cite{heptagon2013}, are respectively compilable in C and Ocaml. Eclat~\cite{sylvestre-eclat} can also be mentioned as a synchronous language for the description of FPGA logical circuits. Such languages provide mathematical properties that make them reliable for the design of (potentially critical) reactive systems. Chips describes systems to make them behave in a similar way. Each component of a Chips application is represented by a functional block with inputs and outputs. Interconnected components executions are synchronized so that the system models some actions as simultaneous. Basic Chips constructs include three kinds of functions:

\begin{itemize}[left=0.5em]
  \item ``pure'', to factorize expressions that may be used elsewhere, in the components' or the system's description (lines 1 to 7 in Listing~\ref{codesample}) or,
  \item ``logical'' if the function models a component of our system, it can include memory in the form of inner variables or execute sequential algorithms (lines 9 to 30 to model the different interacting modules of the application) or,
  \item ``physical'' if the function models a hardware component comprising calculation capacity and allocatable memory (lines 34 and 35 to specify the way a server interfaces the virtual modules and the physical world).
\end{itemize}









\paragraph{Non-standard synchronous language} In contrast to previously mentioned languages, Chips is only partially synchronous. Though logical and physical components are modeled as simultaneously executed, Chips describes the relation between variables in an imperative way, closer to general-purpose programming languages. No parallelism is admitted within the shell of a logical or physical function. Each component's inner variable is modifiable with a C-like syntax. Still, according to synchronous programming principles, the output parameters of a component are only accessed by other components once the whole $Then$ section is completely executed, thus hiding the sequential logic from the whole system point of view. Such constraint ensures the atomicity of the components. Since everything that can be parallelized is set in a different component, the work of splitting apart the different functions among the modeled physical devices is simplified.

\paragraph{Signals aggregation and profiling} An interesting addition to the language is the possibility to specify how a signal can change its cardinality to feed the inputs of a set of components. Directly inspired by AP, Chips provides the \verb'mergeplug' and \verb'splitplug' instructions. When a single signal has to be duplicated to be the input of several components, one can use the \verb'splitplug' construct (see Listing~\ref{codesample2}, line 15). Its first parameter being the name of the signal to be plugged to an arbitrary number of functional blocks. The second one is the name of a function that describes how the signal should be split. Therefore, behind the term \verb'aggregating_function', one can imagine a function that duplicates the signal, that spreads evenly its elements among the plugged components etc. With the same philosophy, the \verb'mergeplug' construct (Listing~\ref{codesample2}, lines 20 and 21) specifies that a signal can be merged with another one according to a certain function, thus allowing to realize summing or filtering operations for instance. 

\paragraph{Refining models with hardware specifications} Additionally to a Chips program, hardware description files should be provided for each physical component (see \verb'import' command in Listing~\ref{codesample}, line 32). Such files would give information about the components capabilities: number of processors, memory size, clock rate, actuators and sensors etc. This additional information enhances the Chips model by allowing to determine at compile time if a program fits a certain hardware or if a signal has to be transmitted to another device for the program to keep running. Syntactically, these constraints are specified in the \verb'SYSTEM' section of a model. Each functional block has to be embedded by a physical device. That dependency is described with the \verb'link' operator (lines 8 to 10 of Listing~\ref{codesample2}).\\

\par These features are what models would be built upon when using Chips. To demonstrate their usefulness, next sections detail their application to the Adaptable TeaStore example. 

\section{Designing an Adaptable TeaStore with Chips}
\label{sec:DesigningAdaptableTeaStore}

In order to construct a solid Model Based implementation of the Adaptable TeaStore, the adopted methodology is the one of CT. One can follow a six steps approach that is decomposed into:
\begin{enumerate}
  \item Identifying \textit{goals}
  \item Identifying \textit{knobs}
  \item Devising the model
  \item Designing the controller
  \item Implementing and integrating the controller
  \item Testing and validating the system.
\end{enumerate}

Section~\ref{subsec:RequirementsAnalysis}, through a rigorous requirement analysis, realizes the steps 1 and 2. The description of the model used for this study is presented in Section~\ref{subsec:TeaStoreChipsModel}. On the base of the built model, Section~\ref{subsec:AchievingAdaptivity} explains the design of the controller used for the system and its integration into our model, thus achieving steps 4 and 5. The last step is kept for the result section of this paper.

\subsection{TeaStore requirement analysis}
\label{subsec:RequirementsAnalysis}

We start by an analysis of the requirements associated with the Adaptable TeaStore case study~\cite{TeaStore}\cite{BDGLZZ25}. When working on the control of a system, two main signals have to be identified, the \textit{goal} and the \textit{knob} signals. 

On the one hand, the \textit{goal} signal is the ideal value the system is aiming for. This could be the response time, the overall CPU load or the memory use of a system. The goal signal serves as input for the controller component introduced by CT to make the system adaptable. 

On the other hand, the \textit{knobs} (or \textit{actuators}) are the signals produced by such controller component. They aim at changing the system's behavior toward desired goals. To identify the knobs, more domain knowledge is required because these signals can sometimes overlap each other or lie in complex parts of the system. For example, if the goal of the controller is to ensure a certain response time, the knobs could correspond to the (de)activation of processor overclocking or to the modification of the available cache size.

\paragraph{Goals identification} Acknowledging common practices of web development, the main goal of the TeaStore application should be to provide an enjoyable online shopping experience. Adaptable TeaStore case study specification~\cite{BDGLZZ25} distinguishes mandatory modules, and optional ones. It could be assumed that the mandatory modules are in charge of the base functionalities the application offers, while the other are additional non-functional requirements to fulfill.

The modules suggested are (mandatory modules in bold font): \textbf{Web User Interface} (WebUI), \textbf{Persistence service}, \textbf{Image providing service}, Authentication functions, recommendation system.
\par From the functional point of view, the TeaStore website should contain images of the articles the store has to sell, suggest items to buy, allow the users to navigate through the pages anonymously and store items in a cart to buy them later. From a non-functional point of view, one could imagine designing an application achieving these goals within a certain time to hold the user's attention, adding authentication services through other websites accounts, having a minimal electrical consumption, actively monitoring the clients behaviors to enhance the suggestion algorithm etc.

Any of these functionalities can be submitted to thorough analysis for a CT application. Our study focuses solely on keeping the response time of the application reasonable as it is a common concern for many websites.

\paragraph{Knobs identification} Now it has been decided what goal to follow, it is possible to identify for each component of the application what are the levers to improve timing performances of the system at execution time. Table~\ref{tableknobs} presents some of the knobs identified for the TeaStore example application, applying control to which could be of interest to reduce delays between operations. For instance, if the user internet connection is unstable, it becomes relevant to turn off image resizing to avoid sending multiple times the same picture if need be to re-send images when packets are lost.

\begin{table}[t]
  \caption{TeaStore Modules and their associated knob for real-time adaptation of services speed}
  \label{tableknobs}
  \tabsize
  \centering
  \settowidth{\tabwidthi}{\thead{Image Provider}}
  \settowidth{\tabwidthii}{Request type (Categorical)}
  \settowidth{\tabwidthiii}{Enable or disable JavaScript if unsupported operation for a browser}
  \begin{tabular}{@{}
      >{\raggedright}p{\tabwidthi}
      >{\raggedright}p{\tabwidthii}
      >{\raggedright}p{\tabwidthiii}
      @{}
    }
    \toprule
    \thead{Module} & \thead{Knob name (and nature)} & \thead{Description}
    \tabularnewline\toprule
    \thead{Web UI}  & JS (Boolean) & Enable or disable JavaScript if unsupported operation for a browser
    \tabularnewline\midrule
    \thead{Image Provider} &
    \multicolumn{2}{l@{}}{%
      \begin{tabular}[m]{@{}
          >{\raggedright}p{\tabwidthii}
          >{\raggedright}p{\tabwidthiii}
          @{}
        }
        Cache size (Natural) & Adjust the number of elements the local cache can store
        \tabularnewline\midrule
        Image size (Natural) & Default size of the images to store in the cache
        \tabularnewline\midrule
        Image resizing (Boolean) & Authorize the resizing of images
      \end{tabular}%
    }%
    \tabularnewline\midrule
    \thead{Persistence} &
    \multicolumn{2}{l@{}}{
      \begin{tabular}[m]{@{}
          >{\raggedright}p{\tabwidthii}
          >{\raggedright}p{\tabwidthiii}
          @{}
        }
        Request type (Categorical) & ``fulldb'',``available'',``batch-queries'', ``top-popularity'', \ldots
        \tabularnewline\midrule
        User profile (Boolean) & Whether to retrieve additional personal data or not
      \end{tabular}%
    }%
    \tabularnewline\midrule
    Authentication &
    \multicolumn{2}{l@{}}{
      \begin{tabular}[m]{@{}
          >{\raggedright}p{\tabwidthii}
          >{\raggedright}p{\tabwidthiii}
          @{}
        }
        Available (Boolean) & Authorize authentication or not
        \tabularnewline\midrule
        Auth kind (Categorical) & ``SSO only'',``TeaStoreAuth only'',``AnyAuth'', ``None'', \ldots
      \end{tabular}%
    }%
    \tabularnewline\midrule
    Recommender & Nb params (Natural) & Number of parameters for the regression algorithm
    \tabularnewline\bottomrule
  \end{tabular}
\end{table}

The experiment conducted in this paper is focused on the development of the \textbf{Image provider}, and, more precisely, on how to use Chips to model a system where load balancing is achieved through the dynamic control of the local cache size. The greatest the number of elements stored in the cache, the lowest the number of requests to the database, and therefore, the less time it takes to answer a user's request.

\subsection{A TeaStore Chips model}
\label{subsec:TeaStoreChipsModel}

\begin{figure}[t]
  \centering
  \includegraphics[width=\columnwidth]{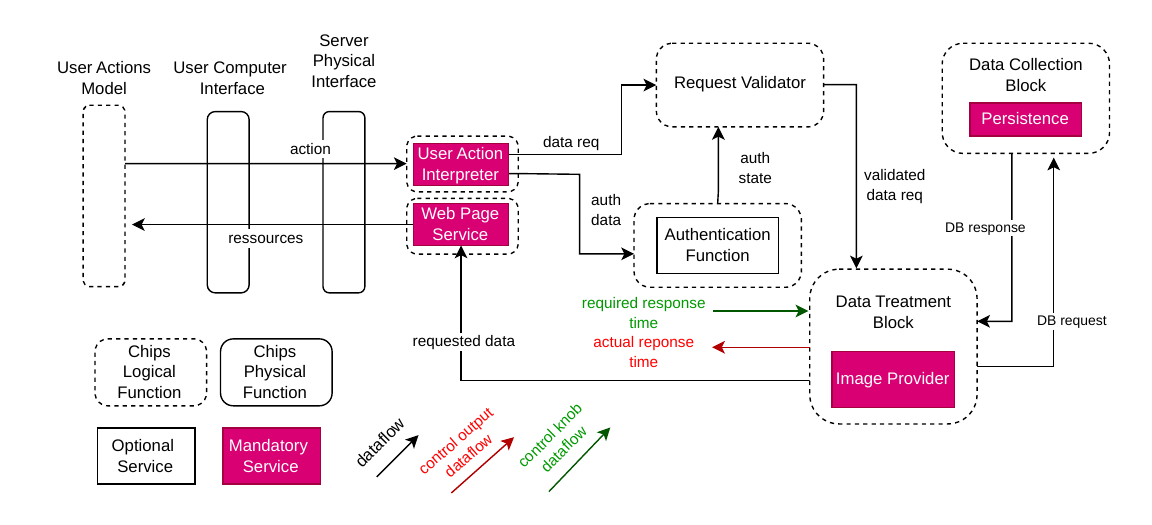}
  \caption{Block diagram of the designed architecture for the TeaStore application and its interaction with a user}\label{generalblockdiagram}
\end{figure}

\paragraph{Model devise} Following CT bases~\cite{hellerstein-feedback-2004}, Chips uses block diagrams. As first step, the system model must be assembled by turning each module of the specifications into a block converting an input signal into an output signal, as illustrated in Figure~\ref{generalblockdiagram}. 
The choice was made in this paper to keep the model as simple as possible. The use case modeled is one where only one user is conversing with the server. Although it is not realistic, it remains easy both to implement and to improve when the need comes to scale the model up.
Other points are worth noticing, there are some differences between the original modules described in Section~\ref{subsec:RequirementsAnalysis} and the diagram presented: 
\begin{itemize}
  \item The WebUI has been split into two components to clearly separate its two functions, receiving the data from the user and sending back a response (respectively \textit{User Action Interpreter} and \textit{Web Page Service}).
  \item The physical interfaces through which data are exchanged are represented (\textit{Server Physical Interface} and \textit{User Computer Interface}).
  \item A \textit{Request Validator} block is there to decide whether a request for data of the user is correct or not according to its authentication state.
\end{itemize}

\begin{figure}
  \centering
  \includegraphics[width=0.8\columnwidth]{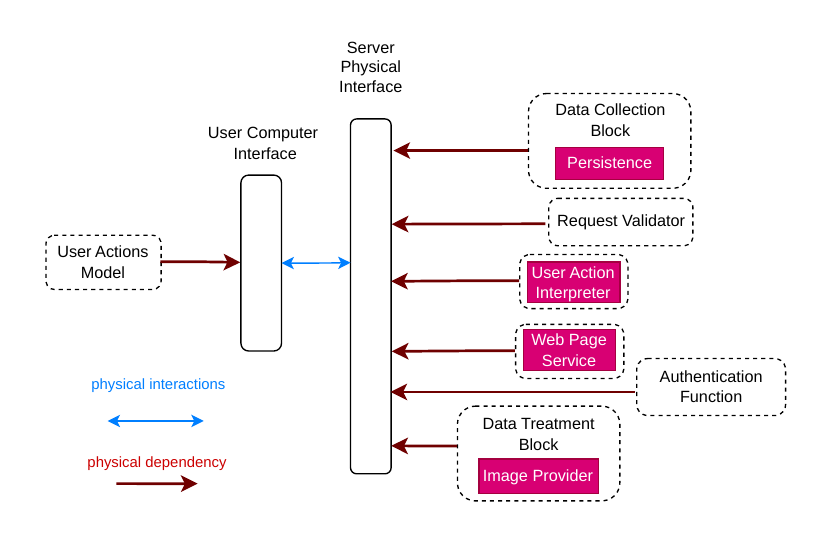}
  \caption{Dependency graph of the designed architecture for the TeaStore application and its interaction with a user}\label{generalphysicaldependency}
\end{figure}

When working on constraints as response time or fault tolerance, it is important to model not only the processed data, but also the context in which it is processed. Hence the differentiation of logical and physical functions of our blocks. In the end, any algorithm will always be executed by a hardware component. This hardware is responsible for the efficiency of the system at a certain task and if we want to keep our model useful, we must take it into account. Figure~\ref{generalphysicaldependency} is complementary to the block diagram described before. The physical dependency graph (associated with Chips \textit{logical} and \textit{physical} keywords and the \textit{link x to y} instruction) allows the model to better determine data transmission timings, calculation speed and memory constraints. When two entities try to communicate while being associated with different physical devices, i.e.,~transmitting a signal through the physical interaction arrow of the diagram, there is a need to encapsulate the information in a protocol. According to the nature and connectivity of the communicating devices, a set of constraints can be applied to keep the model as close to the reality as possible. At simulation time, this could mean virtually adding delays, loosing random packets of data or introducing errors.

To achieve a working model with the aim to parameterize the cache size used by the server, simplifying hypotheses were made:
\begin{itemize}
  \setlength\itemsep{0.3ex}
  \item The user always waits for an answer from the server.
  \item The user never disconnects from the server.
  \item The server provides resources as either \verb'required page' or \verb'authentication page', modeled by a boolean.
  \item No recommendation system is used.
  \item The user always sends actions to the server after receiving resources from it.
  \item Once the user has provided right authentication data, it is considered by the server as \textit{connected} until the end of the simulation.
  \item When the user is connected, the server always responds with the \verb'required page'.
  \item Time delays are negligible, excepts for the ones it takes to fetch images from the cache or from the database.
  \item The server never shuts down and no packets are lost between it and the user computer.
  \item The database always contains the requested data, and provides images one at a time.
  \item The server tries to keep the response time of the image provider around a given constant time value.
  \item User actions are interpreted as a three fields data structure.\\
  ~~\verb'action_request':
  \begin{itemize}[leftmargin=1.5cm] 
    \item[\textbf{bool}] isHeavyRequest
    \item[\textbf{bool}] requestPrivatePage
    \item[\textbf{bool}] providesRightAuthData
  \end{itemize}
\end{itemize}
These assumptions allow abstracting away many constraints that would render the system too complex to analyze. If the control were to be applied to another part of the system, such hypotheses would have to be changed to properly separate the study concerns.

\subsection{Achieving adaptivity}
\label{subsec:AchievingAdaptivity}

\begin{figure}[t]
  \centering
  \includegraphics[width=0.60\columnwidth]{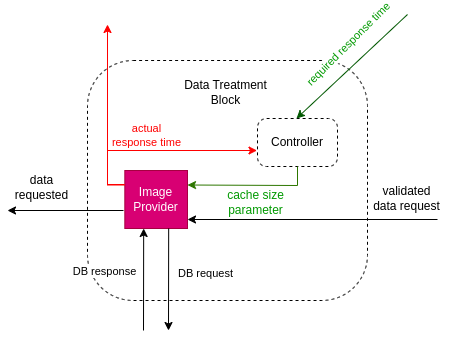}
  \caption{In depth block diagram of the Data Treatment Block}\label{dtblock}
\end{figure}

The component of interest --the cache of our image provider block-- can now be studied without worrying about other sources of disturbances than the one that directly impacts them. To do so, the image provider block includes, on top of the cache, a controller component depicted in Fig.~\ref{dtblock}. 

\paragraph{Controller design} A PID (Proportional Integral Derivative) controller is used to manage the caching part of the TeaStore application. It is a controller that takes into account the error (difference between expected and resulting value) of a system, its integral and its derivative to modify the knob it controls. This kind of controller is used in a vast majority of control scenarios with good results and its simplicity makes it an easy to implement example of controller to work with in our case~\cite{astrom-pid-1995}. 
The inner data treatment block controller will have the task to convert its input command (the response time required by the server general controller) into a cache size (an integer corresponding to the number of items the cache can retain), so that the overall calculation time of the system remains around the command value. The cache used will apply the Least Recently Used strategy (LRU), and to keep the model lightweight, images will be modeled by an ID number between 1 and the total number of images of the database $DB\_SIZE$.
For each request, the image provider will look for the images IDs in the cache. This search is modeled by the $CACHE\_SEARCH\_TIME$ value. For each image not found in the cache, the image provider sends a data request taking $DB\_REQ\_TIME$. Therefore:
\[
actual\_response\_time = CACHE\_SEARCH\_TIME + X \times DB\_REQ\_TIME
\]\[
CACHE\_SIZE(t) = P \times error(t) + I \times \left( \int_0^t error(t) \times dt \right) + D \times \frac{d~error(t)}{dt}
\]
where:
\begin{itemize}
  \item $X$ is the number of images not found in the cache depending on the previous requests and the actual size of the cache,
  \item $P$, $I$ and $D$ are constant real values chosen either empirically, arbitrarily by experience of a CT expert, or analytically by solving the whole system equation,
  \item $error(t) = response\_time\_required - actual\_response\_time$.
\end{itemize}
\vspace{1ex}
\noindent If the cache is too small, $X$ is likely to be high, while if $X \geq DB\_SIZE$, the response time will always be $CACHE\_SEARCH\_TIME$ as the cache can contain the whole database.
To match the behavior of a user who sometimes requires the same images, and sometimes completely different images, the images requested will be randomly generated in $[1,DB\_SIZE]$. That representation could also be upgraded by changing the probability distribution of the chosen numbers to simulate the popularity of different articles.

\section{Extending adaptation to the structure of the system}
\label{sec:APTwist}

\begin{figure}[t]
  \centering
  \includegraphics[width=\columnwidth]{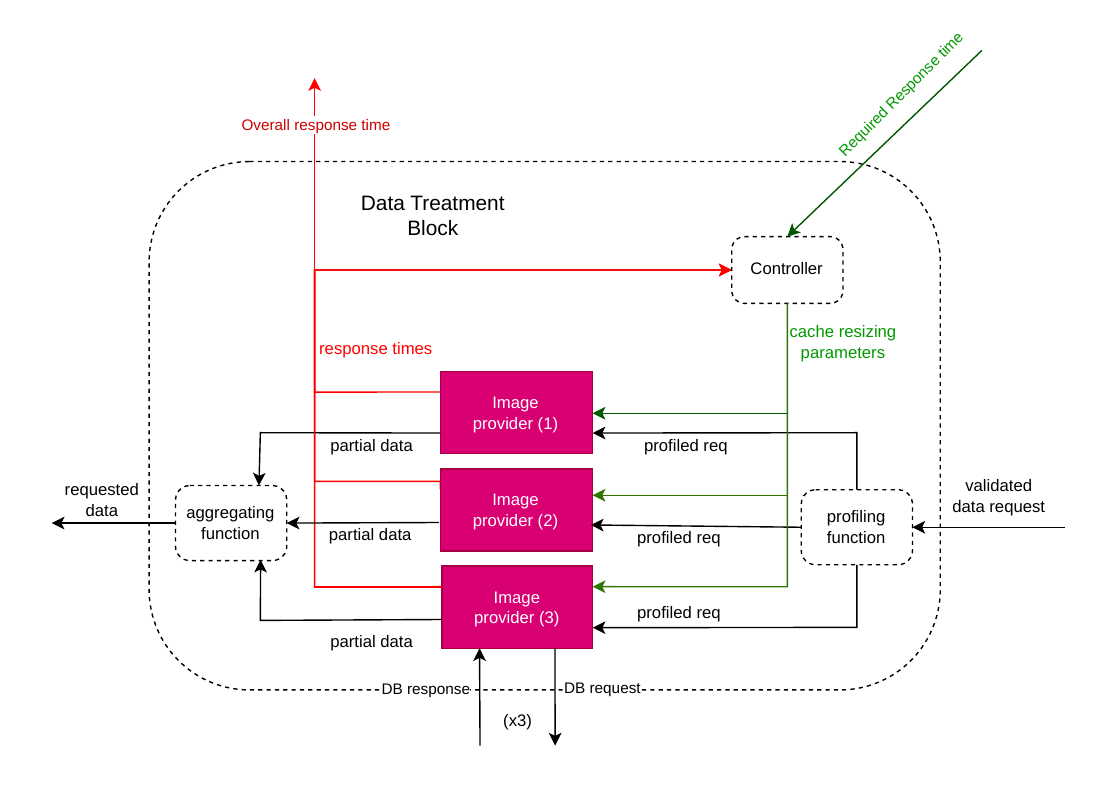}
  \caption{Data Treatment Block handling several Image Provider instances}\label{moreCaches}
\end{figure}

If the running system has multi threading capabilities for parallel computing, one could imagine enhancing the previously described version of the controller (see Figure~\ref{dtblock}) with more instances of image provider (as the example given in the listing~\ref{codesample2} of Section~\ref{sec:ChipsOverview}). In such case, the control logic can remain the same, but the way the command signal is transmitted to the under-control component must be profiled. Such architecture requires to always send the same requests to the same image provider components so the different caches don't store the same information twice. This concept can be expressed in a quite natural way with the use of \verb'mergeplug' and \verb'splitplug' instructions on the outputs and inputs respectively of the image providers working in parallel. They provide the possibility to turn raw signals into an adapted version of it for the image providers to efficiently cooperate(\textit{profiling function} and \textit{aggregating function} in Figure~\ref{moreCaches}).

Thank to these operations, the same encoding of the Data Treatment Block can be used, whatever the number of instances of image provider. It makes it also possible to follow the formalism of~\cite{bliudze:hal-05158948} to build applications with multiple layers of hierarchy for more complex systems. This facility of AP to abstract away complex communication and coordination protocol layers is what Chips is seeking for in term of expressivity of its models. AP makes it easier for a described system to turn the architectural configuration of it into parameters that the model's controllers can tune, hence allowing even more adaptivity. This ``multi-provider'' instance model is provided here as a motivation for Chips further development, but could not be implemented as more work is needed to fully introduce AP syntactic constructions without breaking the rigor the synchronous paradigm offers. This point will be expanded in Section~\ref{sec:future}.

\section{Experimental setup}\label{sec:ExpSetup}

\paragraph{Test and validation of the system} The Chips model for the TeaStore application was manually translated into a BIP set of components according to transformation rules described in Section~\ref{transformation}. The presented experiment was conducted on the BIP model\footnote{The BIP version used is available at~\url{https://gricad-gitlab.univ-grenoble-alpes.fr/verimag/bip/compiler.git}} with a machine using a 13th Gen Intel® Core™ i7\-13850HX processor and a RAM of size 32 Gigabytes. The BIP model for the TeaStore application takes about 2 to 3 minutes to compile into an executable code for simulation. And such simulation takes a few seconds to run with the parameter described in Section~\ref{params}. Each component of the block diagram implemented is modeled by a single BIP component, which contains an automaton having at least as many transitions as the block's inputs and outputs.

\subsection{Chips to BIP transformation}
\label{transformation}
\begin{figure}[t]
  \centering
  \includegraphics[width=\columnwidth]{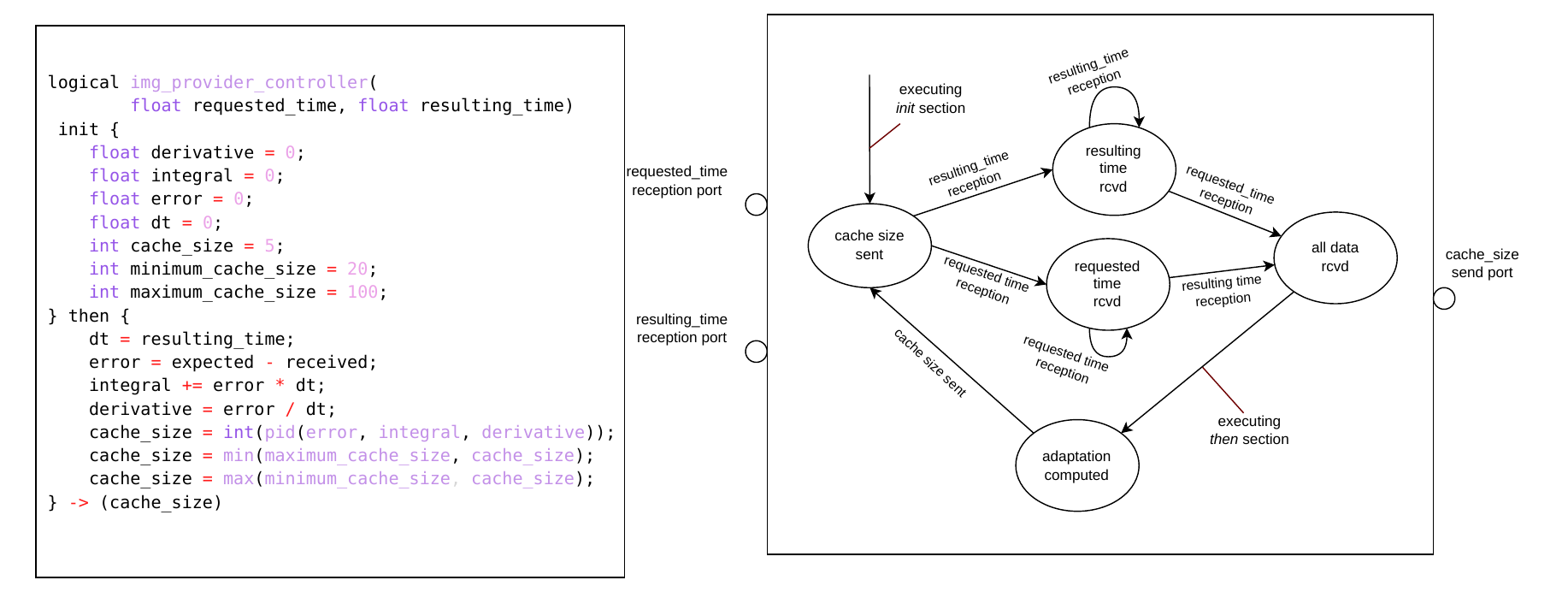}
  \caption{A Chips implementation of a PID controller and its BIP model equivalent}\label{transform}
\end{figure}

Most of the functional blocks of the Chips model described by our model (Figure~\ref{generalblockdiagram}) could be turned into an associated BIP component in the following manner (also shown on Figure~\ref{transform}): 
\begin{itemize}
  \item Each inner variable, input parameter and output parameter of the Chips function is modeled by a variable of the same name in the BIP component.
  \item The automaton associated with a Chips function forms a loop achieving the following three kinds of transitions:
  \begin{itemize}
      \item input transitions synchronized with the output transitions of another component's automaton providing the input data,
      \item updating transitions which are internal (i.e.~not synchronized/exported), to execute $Then$-section operations,
      \item output transitions synchronized with the input transitions of another component's automaton needing data to keep updating.
  \end{itemize}
  \item There is at least one input transition for each input variable of the Chips function.
  \item For each input transition, if not all the input data have been gathered, the transition leads to a state modeling the reception of the data. Such state loops on itself with a transition receiving the same data if the same input is received several times in a row, thus overriding the previous value to only take into account the most recent information.
  \item If the data collected by the input transition is the last remaining input to be collected, the transition leads to an ``all data received'' transition.
  \item An internal transition associated with the $Then$ section is added from ``all data received'' to ``adaptation computed''.
  \item Finally, all the output transitions are added, in any order, each leading to a state modeling the sending of the data. The last one of these states serves as the source of the input transitions. 
\end{itemize}

No final state is required as the system isn't supposed to shut down. The initial state is then set to the state representing the sending of the last output parameter for every component except for an arbitrarily chosen one.

That remaining component will start at the state representing the last input parameter so the system isn't in deadlock. On the initial transition leading the component to its first state, the Chips $init$ section is executed. The only component that did not follow this transformation process is the Data Treatment Block. In Listing~\ref{codesample} line 26, the Chips program doesn't model the interaction with a database component. The BIP implementation that was used for the experiment comprises such interaction, but no Chips primitive can currently express it. Therefore an additional state was added. Indeed, the current semantic of Chips only allows to transmit information from a component to another when the $Then$ section is fully executed (Figure~\ref{BIPAutomatonWithDB}). Sending requests to another component and waiting for its response is currently impossible (or at least, a lot more refactoring has to be done). Designing language constructs to specify inner-$Then$ synchronized operation is one of the current concerns of this research and will be more detailed in Section~\ref{sec:future}.  A graphical representation of the full system transformation result is given in Appendix~\ref{secn:appendix}. 

\subsection{Parameters of the model}
\label{params}

To run this experiment, parameters of the model were arbitrarily chosen. Even if not particularly based on field values or any website response time benchmarks, variables stay relatively coherent in view of assumptions made on the authors' base knowledge on most websites behaviors:
\begin{figure}[t]
  \centering
  \includegraphics[width=0.55\columnwidth]{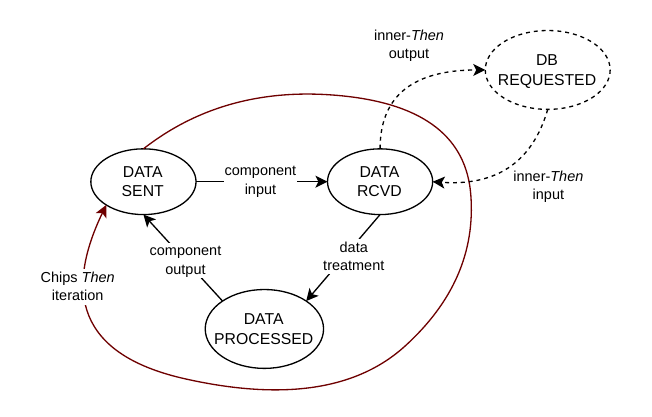}
  \caption{Simplified BIP component for the image provider algorithm (dashed lines represent elements with no Chips equivalent)}\label{BIPAutomatonWithDB}
\end{figure}
\begin{itemize}
  \setlength\itemsep{0.3ex}
  \item The cache search takes less time than the image fetch from the database.
  \item The system aims at reducing the time it takes to provide resources.
  \item The number of images to provide is greater when the user is connected (to provide profile pictures for instance). The presented experiment will not make use of this parameter as it seems to be useful for a more detailed model only. 
  \item The cache has a maximum size (no server has an infinite memory) and a minimum cache size. The minimum cache size was arbitrarily set to 20 as setting a cache size ``too low'' would give an irrelevant hit rate and a website designer would probably better not use any cache than using an almost useless one.
\end{itemize}

Hence the following parameters:
\begin{verbatim}
      float REQUIRED_RESPONSE_TIME = 4.0; 
      float CACHE_SEARCH_TIME = 0.3;
      float DB_REQ_TIME = 2.0;
      float CACHE_MAX_SIZE = 100;
      float CACHE_MIN_SIZE = 20;
      int DB_SIZE = 40;
\end{verbatim}

\subsection{User activity scenario design}
In the follow-up results, the user scenario presented is separated in two phases:
\begin{itemize}
  \setlength\itemsep{0.3ex}
  \item A phase of ``wandering'', where the user model component only sends a request for two images at the same time. It repeats this operation 300 times.
  \item A phase of ``intensive browsing'', where the components send a request for six images and repeat this process until the end of the simulation.
\end{itemize}
Even if this behavior looks more like integration testing than an actual user behavior, it serves well the purpose of this paper as a proof of concept for Chips model design workflow.

\section{Simulation results}
\label{sec:results}

\begin{figure}
  \centering
  \includegraphics[width=0.85\columnwidth]{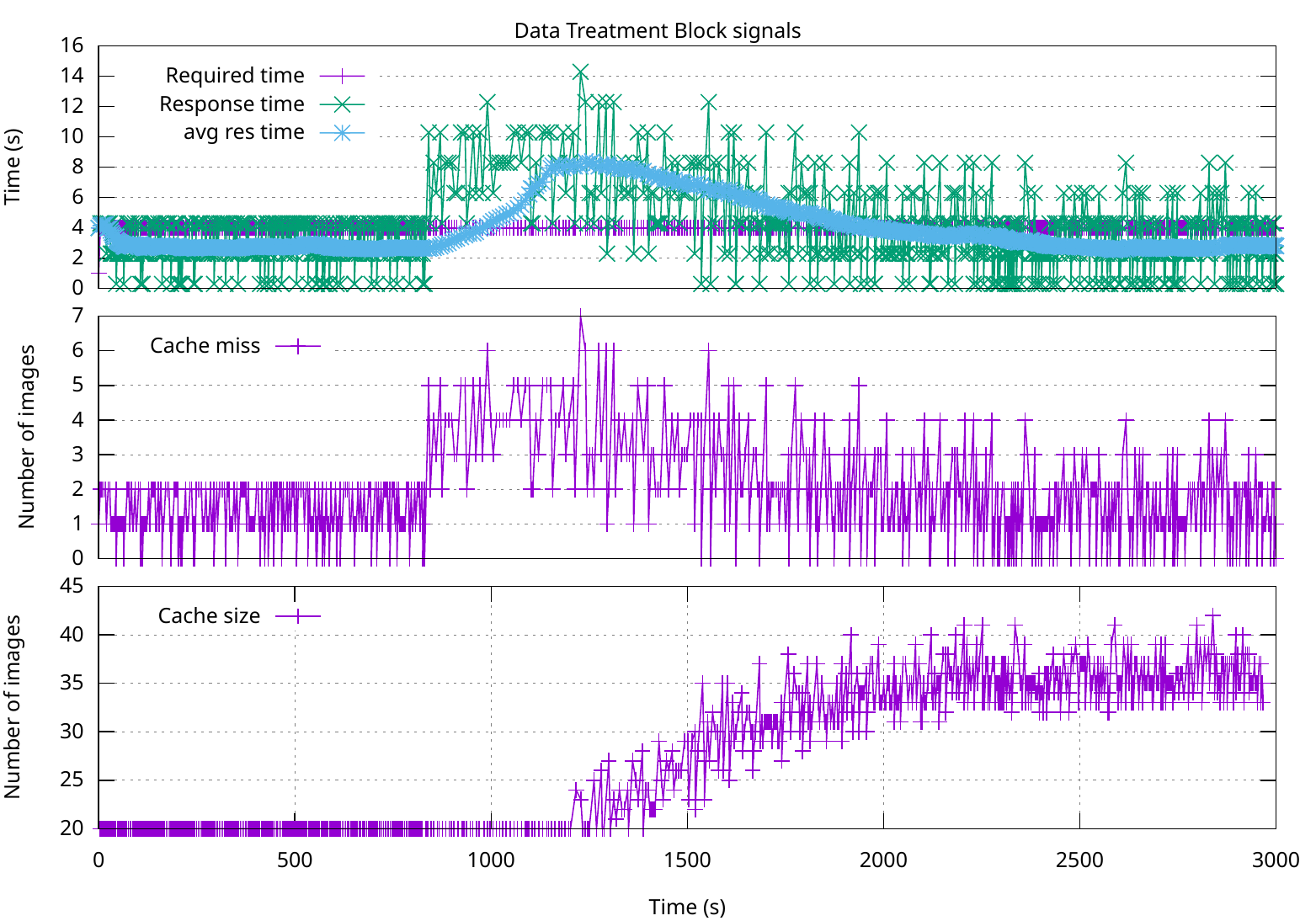}
  \caption{Real time adaptation of the cache size according to the behavior of the client model}\label{results}
\end{figure}

Figure~\ref{results} shows the behavior of the previously described system model. The curves were obtained using the following parameters for our PID controller: $P = -1$, $ I = -0.01$, $ D = 0$. They were empirically tuned until the simulation exhibits signals converging toward a stable state for both the scenario phases. Other design techniques could be used as shown in~\cite{astrom-pid-1995}.

During the first 800 seconds of the simulation, the user requests are answered by the server under 3 seconds on average, which is coherent regarding the number of images requested by the user during their ``wandering phase''. The response time can only be greater than the required response time (4 seconds) when no image is found in the cache (2 cache misses take $2 \times 2.0 + 0.3$ seconds). As half the database can be stored in the cache when it is at its minimum size (20 images), it is likely that images requested by the user can be found, hence the average response time of 2.8 seconds. 

When the user disturbs the system by asking for more images for each request (``intensive browsing phase'') after the 300 requests (t = 834), the response time increases and the controller adapts the cache size to keep the response time around the command. The jittering phenomenon of the cache size around 37 is due to the fact that our system is discrete. It cannot set the response time exactly to the command so it alternates between a response time a little shorter and a little longer than required.

Let us note that by tuning the PID coefficients, it is possible to change the behavior of the system to reach the optimal cache size quicker, at the risk of temporarily setting the cache size way higher than required. It is 40 in this setup since this is the size of the database and having a bigger cache wouldn't give better results.

\section{Related work}
\label{sec:relatedwork}
Formalizing adaptivity to design more robust systems is not new. With Chips, our choice is to exploit the results of CT to leverage the question of the behavioral adaptation. But other formalisms exist. One of the most used models to achieve adaptation in software design is the MAPE-K loop. Firstly introduced in~\cite{kephart-vision-2003}, it describes a systematic way to adapt the behavior of a system by a four-step protocol: Monitoring signals of interest, Analyzing their meaning, Planning a way to adapt the behavior, and Executing such plan by tuning correctly the signals commanding the system. The K stands for the Knowledge on which adaptation strategies lie on. It can be designers' expertise, models of the system or databases of interest for the situation. Compared to CT, MAPE-K provides a wider framework to describe adaptation strategies. It has a qualitative approach to the way systems work when CT is more about finding the signals interacting together and formally describe their relation to deduce the best adaptation function.

Still, MAPE-K stands as the base of a lot of research around adaptivity. For example, MAPE-K gave birth to Rainbow~\cite{huang-rainbow-2004}, a framework using a model of the controlled system at run-time to deduce best adaptation strategies at the whole system level. The MAPE-K methodology is then applied to the distributed system under control. However, the development of such adaptation mechanism remains monolithic. It is therefore inherently hard to upgrade and pushes toward a centralized implementation, potentially acting as performance bottleneck.

Aspect Oriented Programming (AOP)~\cite{haesevoets-weaving-2010} proposes to overcome the adaptation workload distribution problem by providing tools for the task distribution at the language level. Though it comes with other challenges like the need for synchronization of wide scale systems. Chips distributes the control in a similar way: the language provides an easy way to reconfigure the models to try other distributions of the tasks. Therefore, it currently suffers the same struggle.

Another problem to overcome with adaptation is the one of anticipation: At design time, it is only possible to anticipate a subset of all the events that could lead to unintended behaviors. One may need to improve the way adaptation is performed while the system is running, when the unwanted event occurs. To do this, the language AIOCJ~\cite{dalla-preda-aiocj-2014} presents a way to modify the code of an application at run-time and to allocate that code to components assuming a certain role in the application.

In the recent years, the development of micro-services applications became more and more popular. Their Application Programming Interfaces (API) allow to think of adaptation, not as a software grafted to an architecture to adapt, but rather as a service to which components of the system can subscribe. This is the approach adopted by Adaptiflow~\cite{ZBQ25}, a java framework that allows exporting the load of the MAPE-K loop to a third-party server at the cost of depending on the availability of the service.

Though, MAPE-K is not the only model that can be used to achieve some sort of real-time adaptation. It can even be considered as too reductive to embrace the complexity of some softwares~\cite{de-lemos-software-2023} and trying to find ways to emancipate from it could lead to even more efficient solutions. With the recent popularity gained by the field of artificial intelligence, the AWARE model~\cite{sanwouo-breaking-2025} has been developed to improve the reflexivity of the feedback mechanism by introducing intelligent agents to the core of the adaptation mechanism. These agents learn the behavior of the system under control while it is running.

Since Chips primary focus is on CPSs, it is dealing with devices often offering low resources in term of memory, communication or CPU. Therefore, run-time code modification, delegation to micro-services or inclusion of AI agents aren't a priority.

\section{Future work}
\label{sec:future}

As Chips is still in its early days, most of its development stack is yet to create. Many features are envisioned to make this project applicable. Following paragraphs detail ongoing work on the language.

\paragraph{A complete version of the compiler} Currently, the language has a syntax that can be parsed, but not interpreted. A meta-model of the language is defined for the main Chips constructs but no compiler exist for the language yet. The hardware description file structure is not fully determined either. These elements are the current priorities of the Chips project and shall soon be developed. Our intention is to use ATL~\cite{jouault-atl-2008} or another model-to-model transformation tool to turn the Chips meta-model into the BIP meta-model, thereof, to generate a robust Chips-to-BIP compiler. Until then, the experimentation presented here could only be realized after a hand-written compilation.

\paragraph{New language constructs} Few models were built thank to Chips, and as the language will be applied to more case studies, new needs are expected to emerge in terms of the Chips expressivity. Some interesting features have already appeared though:

\begin{itemize}
  \item \textbf{Building AP primitives}: The specification of the \verb'mergeplug' and \verb'splitplug' is currently incomplete. More formalism is required for a proper description of what properties the profiling/aggregating functions should have. However, BIP connectors simulating the concept of AP ``arity-modification'' instructions are already under development. Their linked list version is currently functional, but it couldn't be added to our model to exhibit significant results in time for the edition of this document. We intend to design a tree-like version of such connectors to fit a wider range of configurations the different modeled modules can form. 
  \item \textbf{Inner-\textit{Then} data synchronization}: As mentioned in the last paragraph of Section~\ref{transformation}, Chips lacks a way to signify a datum is to be read or written during the update of the variables of a component. A possibility would be to introduce a subtype for the inner variables of a component to turn them into some sort of public attributes in the style of object oriented programming. Such decision isn't made yet and has to be confronted with the formalism Chips tries to adopt so the language doesn't become too permissive and prone to errors.
  \item \textbf{Hierarchical control features}: With its building and plugging blocks design, Chips also permits the description of hierarchically organized control. Adding layers of control for a complex system could allow each sub-system to work at its own pace in a correct way, while together reaching a larger goal~\cite{scattolini-architectures-2009}. This layering could be achieved by adding annotations to our Chips blocks for instance. It would explicitly show the encapsulation of signals within a control layer.
  \item \textbf{Introducing GALS mechanisms(Globally Asynchronous Locally Synchronous)}: Currently, Chips describes complex systems in the form of synchronous programs. The synchronous hypothesis is very convenient but can't hold if the models described get too large. Transmission delays and calculation time can be considered instantaneous as long as they do not add up too much. Applying GALS knowledge and formalisms~\cite{zhang-system-level-2025} to enhance Chips would open the way for better up-scaling of the models designed. 
  \item \textbf{Virtual clock system}: Finally, Chips models relevance should as well be ensured at the level of the timing constraints of the physical devices they represent. Clocks of a multiple devices system should be handled automatically thank to the use of the hardware description files. The BIP clock component would only serve as a virtual helper for simulation and model checking. It shall be automatically removed at the lowest level compilation of the programs when implementing models on real systems.
\end{itemize}

\section{Conclusion}\label{secn:concl}

This paper described a model of adaptable cloud architecture for a variation of the TeaStore web application that has been implemented. The obtained results showed that control theory can effectively be applied to such computer system to improve its performances. Its design followed a full workflow:
\begin{itemize}[left=1em]
  \setlength\itemsep{0.1ex}
  \item the rigorous analysis of the web application goal (keeping the server response time low),
  \item the distinction of the different means of actions to tune the system in order to achieve these goals (modifying the size of the cache at runtime),
  \item the design of a control theory block diagram of the system directly translatable to a Chips model (Figure~\ref{generalblockdiagram}),
  \item the design of a controller within the Chips model for a chosen component (Figure~\ref{dtblock}),
  \item the systematic transformation of the Chips model to a BIP one,
  \item the validation of the system design by simulating the execution of the BIP model and fine-tuning its parameters.
\end{itemize}

During the development of this sample project, simplifying decisions were made at different levels. The first ones were about the description of the model where the recommendation block was removed and only one user was modeled. Then, the real-time optimization of the cache size was chosen for this study, thus orienting the choice of the information held by the signals exchanged between the user and the server models. Finally, the kind of controller used was one of the most classical that exists to keep the demonstration as generic and easy to repeat as possible. According to the designer's will and necessities, other choices can be made to better suit a different project.

\bibliographystyle{eptcs}
\bibliography{refs}

\appendix

\section{Appendix: Additional figures}
\label{secn:appendix}

\begin{figure}
    \centering
    \includegraphics[width=\textwidth]{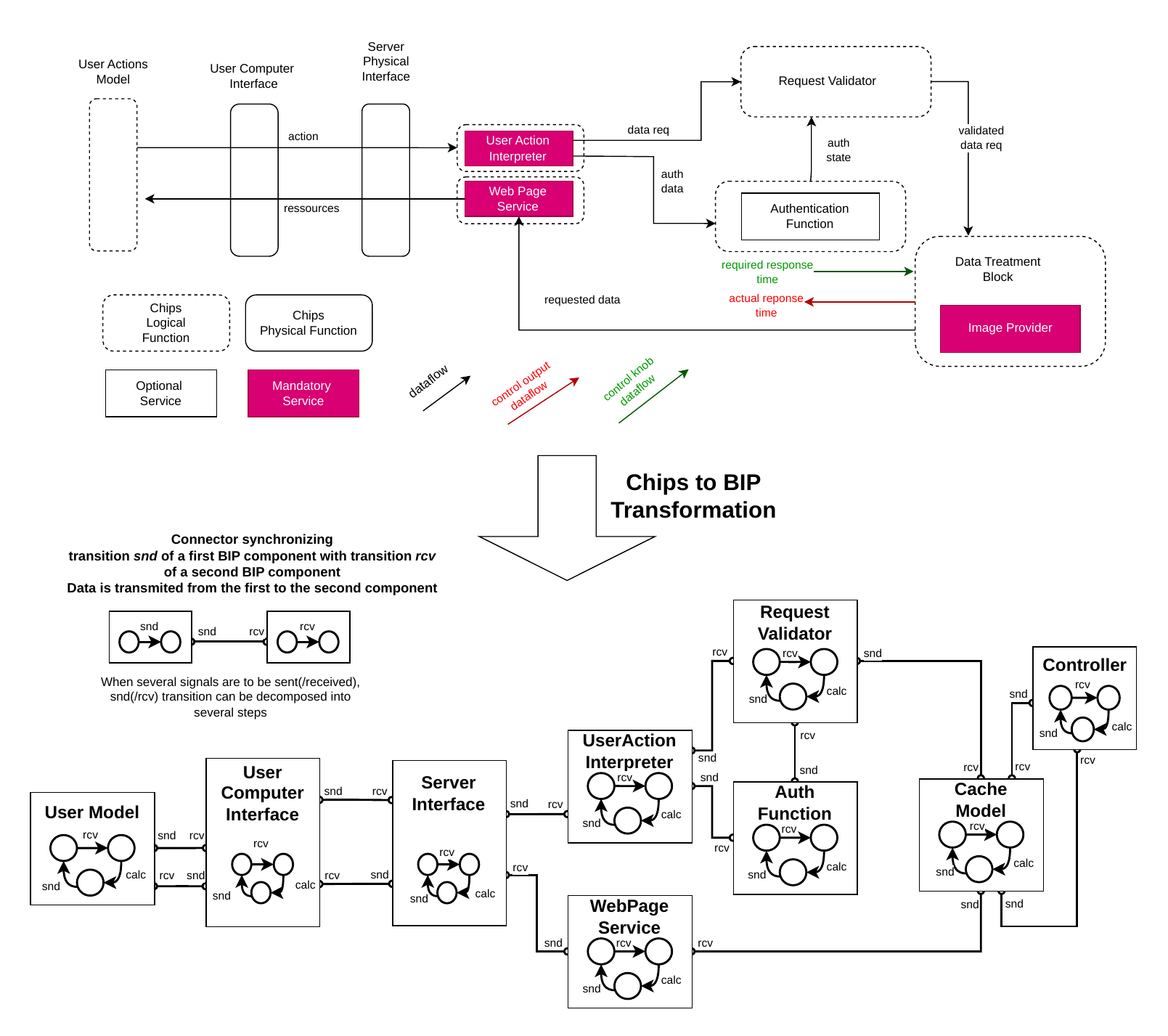}
    \caption{General overview of the full transformation of the Chips Adaptable TeaStore specification into a set of interacting BIP components (Each Chips functional block is turned into a BIP component synchronizing the automaton it holds with the automata of other connected BIP components.)}
    \label{fig:fullTransformation}
\end{figure}

\begin{figure}
    \centering
    \includegraphics[scale=0.75]{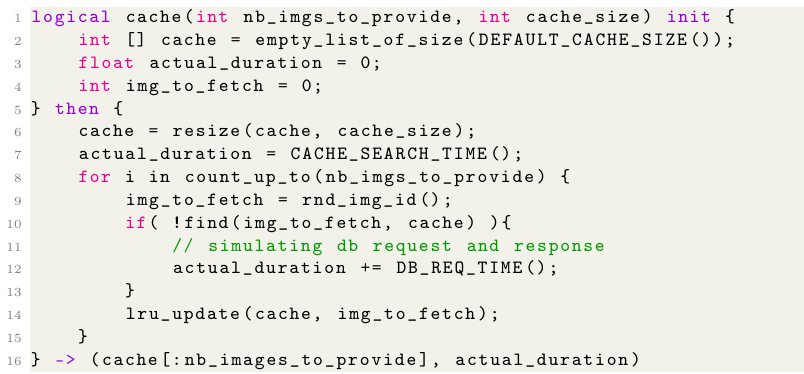}
    
    \includegraphics[width=0.49\textwidth]{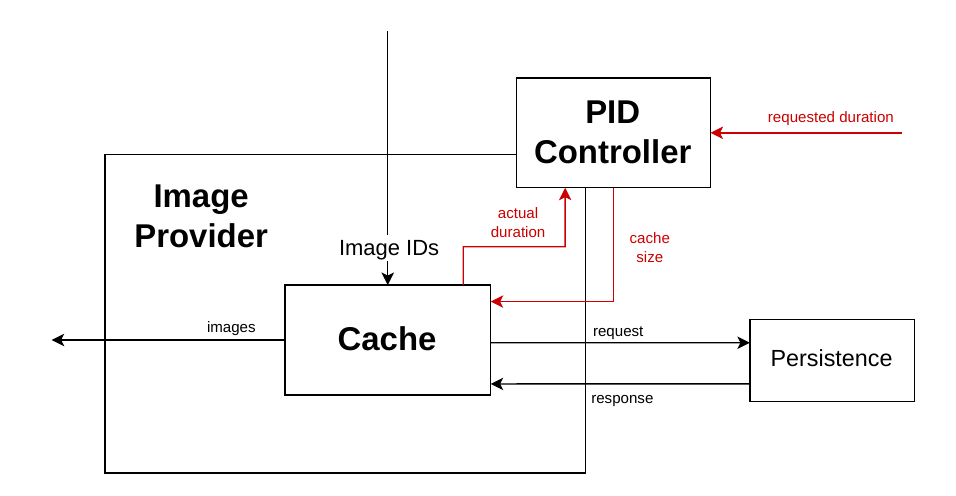}
    \caption{Chips Cache specification and its integration into the Image Provider component of the Adaptable TeaStore (Here the persistence functional block is simply modeled by the addition of a delay in the cache function according to the number of missing images.)}
    \label{fig:cacheModel}
\end{figure}

\begin{figure}
    \centering
    \includegraphics[scale=0.75]{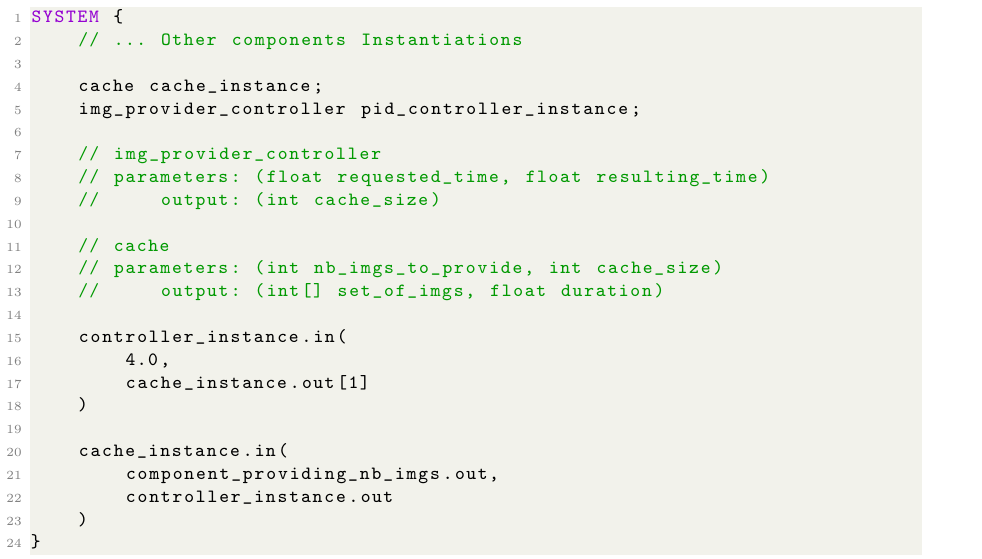}
    \caption{Chips instantiation and connection of the model elements presented in Figure~\ref{fig:cacheModel} (The PID controller corresponds to the component modeled in Figure~\ref{transform}.)}
    \label{fig:Instantiation}
\end{figure}

Below, we provide three additional figures to allow the reader to
\begin{itemize}
    \item Apprehend the scope of a Chips full design of an application model, namely, the Adaptable TeaStore (Figure~\ref{fig:fullTransformation});
    \item Better understand the way a component/functional block can be defined (Figure~\ref{fig:cacheModel}); 
    \item Clearly distinguish how such objects get instantiated in the context of a wider model (Figure~\ref{fig:Instantiation}).
\end{itemize}

\end{document}